# The detectability of planetary companions of compact galactic objects from their effects on microlensed lightcurves of distant stars

Alberto D. Bolatto

Astronomy Department, Boston University

725 Commonwealth Avenue, Boston MA 02115

and

Universidad de la República, Facultad de Ciencias

Tristán Narvaja 1674, Montevideo, Uruguay

and

Emilio E. Falco

Harvard-Smithsonian Center for Astrophysics

60 Garden Street, Cambridge MA 02138

## ABSTRACT

We discuss a possible method for detection of dark companions of galactic objects of stellar mass. Such binary systems are likely to occur in the galactic disk and possibly also in the halo. The high incidence of binary and higher-multiplicity systems in the solar neighborhood, if indicative of the galactic disk at large, implies that current searches for the gravitational microlensing signature of massive compact objects in our galaxy would yield a significant fraction of binary systems. Our calculations suggest that 40% of the lightcurves that will be obtained in such searches may be sufficiently perturbed to reveal, if sufficiently well-sampled, the presence of a compact dark companion of Jovian mass orbiting a primary. The likelihood of occurrence of perturbed lightcurves must also be taken into account by systematic search programs, to improve their event detection efficiency. The statistics of perturbed lensing events, if carefully interpreted, could yield estimates of the incidence of binary systems with low mass ratios, including that of systems with planets.

*Subject headings:* dark matter - Galaxy: halo - gravitational lensing - planetary systems - stars: low-mass, brown dwarfs

## 1. Introduction

Recent observations (Alcock et al. 1993, Aubourg et al. 1993, Udalski et al. 1993, Udalski et al. 1994) reveal that dark matter in our galaxy may comprise a significant population of massive compact halo or disk objects (hereafter, for brevity we refer to both varieties as MACHOs). The detection of such objects was accomplished through measurements of the variability of stars in the LMC and in the galactic bulge, induced by the gravitational lens effect of the MACHOs. The magnification varies measurably with time due to the relative motions of source, lens and observer. Single galactic lensing objects yield lightcurves with simple Lorentzian profiles, amplitudes of up to several magnitudes, and timescales of hours to weeks, for stellar masses and for typical expected velocities in our galaxy. Binary lensing objects yield lightcurves that differ significantly from those for single (point-like) objects. Such differences were exploited as a possible method for detecting compact low-mass companions of stars and of other compact objects in the galactic disk (Mao & Paczyński 1991, hereafter MP; Gould & Loeb 1992, hereafter GL). The approach of MP was extended in GL, to take better account of the effects of passage through the effective region of influence of microlensing binaries. We have studied a simple model that reveals a relatively high probability of detection of planetary systems, even somewhat higher than was estimated in GL.

We describe our model in §2, our results in §3. We describe briefly the current state of observations of galactic microlensing events in §4. Finally, we conclude and discuss the significance of our results in §5.

## 2. Model

For our calculations, we adopted the MP model: the lensed distant source is a star, and the lensing system contains two objects of sizes ranging from planetary to stellar (we refer to such systems as binary lenses). The primary object is assumed to be a star with mass $M_1 = 1 M_\odot$; the mass of the secondary is $M_2 = qM_1$. We consider mass ratios $q = 0.01$ and $q = 0.001$; these values suffice to demonstrate the effects of a family of secondaries, ranging from planets to brown dwarfs. For a typical stellar radius $R \sim 10^{11}$cm, and $R \ll q^{1/2} R_0 D_s / D_d$, where $R_0$ is the Einstein radius (e.g., Paczyński 1991) for a 1 $M_\odot$ point-like mass, and $D_s$ (range: 10 − 60 kpc for stars in the bulge or in the LMC) and $D_d$ (range: 5 − 20 kpc for objects in the disk or in the halo of the Galaxy) are the distances to the lensed star and to the deflecting object, respectively. Therefore, we safely assume that the source is point-like in all our calculations. (Such an approximation breaks down for planetary companions of terrestrial mass.)

The relative motions of a lensing point mass and a distant source produce characteristic lightcurves with Lorentzian profiles; see e.g., Paczyński 1986 and Alcock et al. 1993 for examples of theoretical and measured lightcurves. An orbiting companion can significantly perturb the gravitational potential of the lens, and thereby appreciably perturb the lightcurves. The shapes of lightcurves for binary lenses depend solely on four parameters: the mass ratio $q$, the separation between the binary components (projected on the sky) $a$, the impact parameter $p$ (with respect to the primary) and the impact angle $\phi$. The first two parameters determine the mapping from the lens plane to the source plane, and therefore the topology of the configuration. If the perturbations of lightcurves are measurable, observations of microlensing events become a useful scheme for detecting dark companions (GL).

In the approach of MP, strong binary microlensing is characterized by the surface area (cross-section) enclosed in the caustic curves of the lens in the source plane (Witt 1990). When the source finds itself within such a region, five images appear (rather than three as in the normal case), with a corresponding increase in the magnification in the form of sharp peaks when the lensed star is unresolved. Two measures of the cross-section were proposed in MP: the width $\Delta p$, the range of impact parameters for which the source lies within the area surrounded by the caustics, averaged over all impact angles, and the length $\Delta l$, the length of a segment of the trajectory of the source that lies within the caustics, averaged over all impact angles and parameters. The mean width is $\Delta p$ averaged over the distribution of separations, $a$, of the binaries. Known binary stars in the galactic disk are distributed approximately uniformly in the logarithm of their period, $\log P_{bin}$ (Abt 1983), such that roughly 10% are found in each decade of period. We calculate the mean "width" as a function of $q$ (see Eq. 4 of MP); we assume as in MP that the distribution applies to the binary systems in our model. (For halo objects, binary formation mechanisms could easily differ from those operating in the galactic disk.) For such systems, assuming that the secondary orbits the primary at a distance 1 AU $\lesssim a \lesssim$ 10 AU, the corresponding width in units of $R_0$ may be estimated as in MP:

$$w_{pl} = \int_{-0.5}^{0.5} \Delta p \quad d\log(a/R_0). \tag{1}$$

The integration limits correspond to microlensing of stars in the bulge, but they also apply to halo stars if we are not concerned about slightly underestimating $w_{pl}$ for such stars.

The calculations of MP and GL, within approximations similar to those stated above, showed that 5 to 20% of the lightcurves could be unambiguously attributed to binary systems, where the companions have planetary (roughly Jovian) masses. In MP, the authors pointed out that it is not necessary for the source to actually penetrate the region surrounded by the caustics to obtain a significantly perturbed lightcurve. The significance of the effect is illustrated in figure 1. The observable consequences of passage near caustics were not calculated in detail in MP (see also GL). Thus, we were led to a reformulation of the width and length of MP for a binary lens. We wished to quantify the difference between the lightcurve produced by a binary lens and that produced by a single point mass. We adopted the following definitions:

**width** $\Delta p$  *The range of impact parameters for which the distance (the absolute value of the difference) between the lightcurves due to the binary lens and to the point-mass lens exceeds a certain threshold, averaged over all impact angles $\phi$.*

**event length** $\Delta t$  *The length of time during which the ordinates of the curves differ by more than a certain threshold, averaged over all impact angles $\phi$ and impact parameters $p$.*

We chose a global measure for our purposes: *the distance between two functions is the integral (area) of the absolute value of the difference between the two functions.* Excursions above and below the lightcurve due to a single point mass then contribute to such a distance. Our strategy was:

1. Compute a series of lightcurves for topology $(q, a)$ and fixed impact parameter $p$, for a range of impact angle $\phi$.

2. Compute the lightcurve for a point-mass lens with impact parameter $p$. Such a curve asymptotically merges with the wings of the curves due to binary microlensing.

3. Integrate the absolute value of the difference between the curves calculated in 1 and 2 for each impact angle, then average over $\phi$ assuming a uniform distribution. Thus, we obtain $\sigma$, the mean distance between single and binary lightcurves.

4. Repeat steps 1–3 for a range of impact parameters $p$. A threshold value of $\sigma$ may now be chosen, yielding a corresponding value of $p$. That value is what we now call "width", $\Delta p_{th} = \Delta p(\sigma_{th})$.

5. Apply Eq. 1, thus obtaining $w(q, \sigma_{th})$.

## 3. Results

For each value of $q$ that we selected, we covered the parameter space with 15 uniform logarithmic steps in $a$ centered on $a = 1$, 7 uniform logarithmic steps in $p$ betweeen $-1$ and 0.2, and 18 steps of 10° in $\phi$ between 5° and 175° (taking advantage of the reflection symmetry about the radius vector from the primary to the secondary). We found that such a grid yielded sufficient accuracy and coverage of the relevant parameter ranges. We applied two algorithms, grid-search followed by Newton-Raphson transport (see, e.g., Schneider, Ehlers & Falco 1992), to determine the positions and magnifications of images for a uniform distribution of source positions along straight tracks.

As an illustration, in figure 2 we plot the widths, averaged over a uniform distribution of impact angle, for $q = 0.001$. To exploit these curves, a detectability threshold is needed. We define one as the product of a minimum detectable magnitude difference and a time span of observations, to ensure that a lightcurve perturbation is not a spurious effect. One of the first projects to announce a detection of MACHOs monitors $\sim 2 \times 10^6$ stars (Alcock et al. 1993). The uncertainties of the photometry, mainly due to local weather and seeing, are between 6 and 10%, corresponding to brightness thresholds for detection in the range of 0.06-0.10 magnitudes. To obtain conservative limits, let us consider sources in the LMC, $R_0 \simeq 1.2 \times 10^{14}$ cm. Following Griest (1991), for a typical tangential velocity $v_T$ in the halo (with stationary sources and observer) of 200 km s$^{-1}$, we can compute the dimensionless time $\tau = v_T \times t/R_0 \simeq 0.01 \, t$ with $t$ expressed in days. A detection threshold of $2 \times 10^{-3}$ corresponds to detecting a deviation of 0.1 magnitudes over 36 hours, or of 0.05 magnitudes over 3 days, in accordance with the timescale of perturbations due to a companion of mass $\sim 10^{-3} M_\odot$. Similarly, if we consider sources in the galactic bulge, $R_0 \simeq 6.7 \times 10^{13}$, $v_T \sim 150$ km s$^{-1}$ and $\tau \simeq 0.015 \, t$ with a timescale for $10^{-3} M_\odot$ of about one day (Griest et al. 1991). We set the threshold at 0.01 (0.002) for $q = 0.01$ ($q = 0.001$), to approximate the expected observational efficiencies.

We calculated the mean "width" $\Delta p_{th}(a/R_0)$ for each value of the threshold, as described above (essentially from figure 2); by numerical integration of Eq. 1,

$$w_{pl} = \int_{-0.5}^{0.5} \Delta p_{th} d\log(a/R_0) \simeq 0.42 R_0 \qquad (2)$$

for a Jovian companion ($q = 0.001$) and

$$w_{pl} = \int_{-0.5}^{0.5} \Delta p_{th} d\log(a/R_0) \simeq 0.95 R_0 \qquad (3)$$

for a brown dwarf companion ($q = 0.01$).

An estimate of the cross-section for a single point mass is $R_0$ (for consistency with our interpretation of *width*, we use $R_0$ instead of $2R_0$ as in MP). Thus, if all stars and compact objects in the galactic disk were in binary systems, our results show that 40% (95 %) of all measured lightcurves could be identified as produced by binary systems with companions of mass ratio $q = 0.001$ ($q = 0.01$) in an observational program such as that of Udalski et al. (1993), and possibly that of Alcock et al. (1993).

## 4. State of the Observations

Recently, Alcock et al. 1993, Aubourg et al. 1993, and Udalski et al. 1993, 1994 reported the first likely detections of a total of 7 events due to MACHOs in our galaxy. It is too early to say whether the expected rates of detection will be confirmed, and whether or not the events themselves will survive as bonafide microlensing events, or prove to be a new type of stellar variability. Since it appears that binary microlensing events may be relatively plentiful, search programs that base their detection schemes on the simplicity and symmetry of microlensing lightcurves should examine their results carefully. Since the effect of a companion grows stronger as the source approaches the primary, the central portions of lightcurves are the most perturbed. Therefore, when fitting a (one point-mass) theoretical curve to the observations, the data in the wings of the lightcurves (which should be least perturbed and hence the smoothest and most reliable parts of the curves) are of crucial importance. Another effect to be taken into account when calculating the statistics of MACHO detections is that heavily perturbed lightcurves, as those for $q = 1 - 0.1$, will be more easily discarded when trying to identify the microlensing events based on their symmetry and simplicity. One way to overcome this problem is to confirm and utilize the achromaticity of lightcurves when trying to detect microlensing-induced fluctuations.

Statistics of the number of perturbed microlensing events could lead to a determination of the incidence of binaries in the disk/halo populations. The fact that the effect of objects with planetary mass is not negligible may also allow estimates of the frequency of occurrence of low-mass dark companions.

The outlying point near the maximum of the lightcurve reported by Alcock et al. (1993) might be explained as a perturbation caused by a dark companion. Unfortunately, due to the paucity of the observations near the peak, a more detailed analysis is not feasible.

## 5. Conclusions

Little is known about the formation history of compact objects in our own and in other galaxies. Based on the knowledge of the high incidence of binaries in the vicinity of the solar system (Abt 1983; Trimble 1990), we have assumed that all compact objects are in binary systems with low $q$. Under such an assumption, and closely following the framework laid out in MP, we have confirmed that microlensing binary systems produce lightcurves that differ in a significant and measurable fashion from those due to single lensing objects. The results of our calculations of approximate detection probabilities, 40% for a Jovian-mass companion and higher than 95% for a low-mass brown dwarf, present encouraging prospects.

As we discussed in §2 and §3, our thresholds were selected to reflect realistic observing conditions, and to yield lower limits for the cross-sections. Thus, our cross-section estimates are significantly larger than those in MP, mainly because we explicitly take into account the effect of passage near, but not necessarily through caustic curves. Our estimates are also somewhat larger than those in GL, because our definition of cross-section takes direct account of the two critical observed parameters for perturbed lightcurves: magnitude difference and event length.

Our calculations are directly relevant to the lensing by matter in the galactic disk. If compact objects in the halo had the same incidence of binary systems as those in the galactic disk, one could speculate that our results are also valid for the halo. Our results are independent of the scale set by the mass of the primary $M_1^{1/2}$, within the bounds of applicability of our calculations. The lower bound for the mass of a planetary companion in this context is about one earth mass.

The detection of dark companions using the microlensing effect suffers from the problem of small amplitudes common to other methods of detection (see, e.g., Tutukov 1987), but it is not affected by the orientation of the plane of the orbits. One significant problem of the microlensing method is the burdensome requirement of a high density of coverage of lightcurves, which imposes a very large data load. However, the experience of three different groups has confirmed the feasibility of such projects.

The recent likely detections of MACHOs in the halo and in the disk of our Galaxy provide a fresh incentive to search for possible binary events in the myriad lightcurves that are becoming available. Furthermore, the high probability of occurrence of distorted lightcurves must be taken into account when observed lightcurves are analyzed, and when detection thresholds are adopted.

We thank PEDECIBA (Programa de Desarrollo de Ciencias Básicas, Uruguay) for support that assisted us in the pursuit of our work on this subject. We also thank S. Mao and J. Wambsganss for useful comments, J. Titilah for assistance with the manuscript, and an anonymous referee for pointing out a potentially significant numerical error in our manuscript.

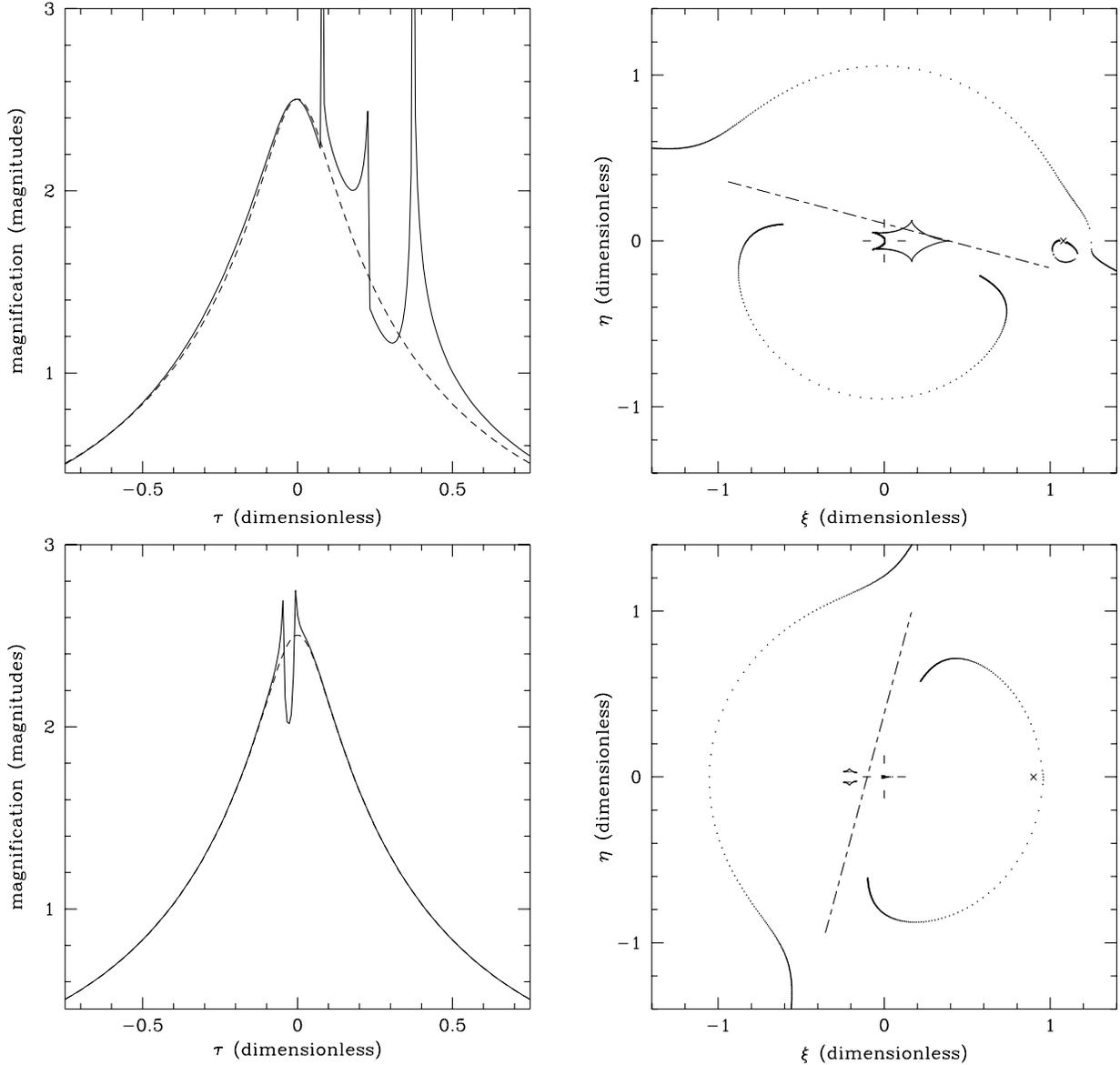

Fig. 1.— Lightcurves (left) and the corresponding trajectories of the images (right), for mass ratios $q = 0.01$ (top) and $q = 0.001$ (bottom) showing that a binary signature is apparent in the lightcurve, even if the source does not cross a caustic as occurs in the $q = 0.001$ case. The position of the primary (secondary) is indicated by a cross-hair (cross). The lightcurve that would be produced by a single point mass is the dotted curve that tracks along the solid curve that corresponds to the binary lens. The unperturbed trajectory of the source is shown (dash-dotted), and the corresponding images are shown as a succession of solid triangles. The curves with cusps near the center of each panel (right) are the caustics for each lens system.

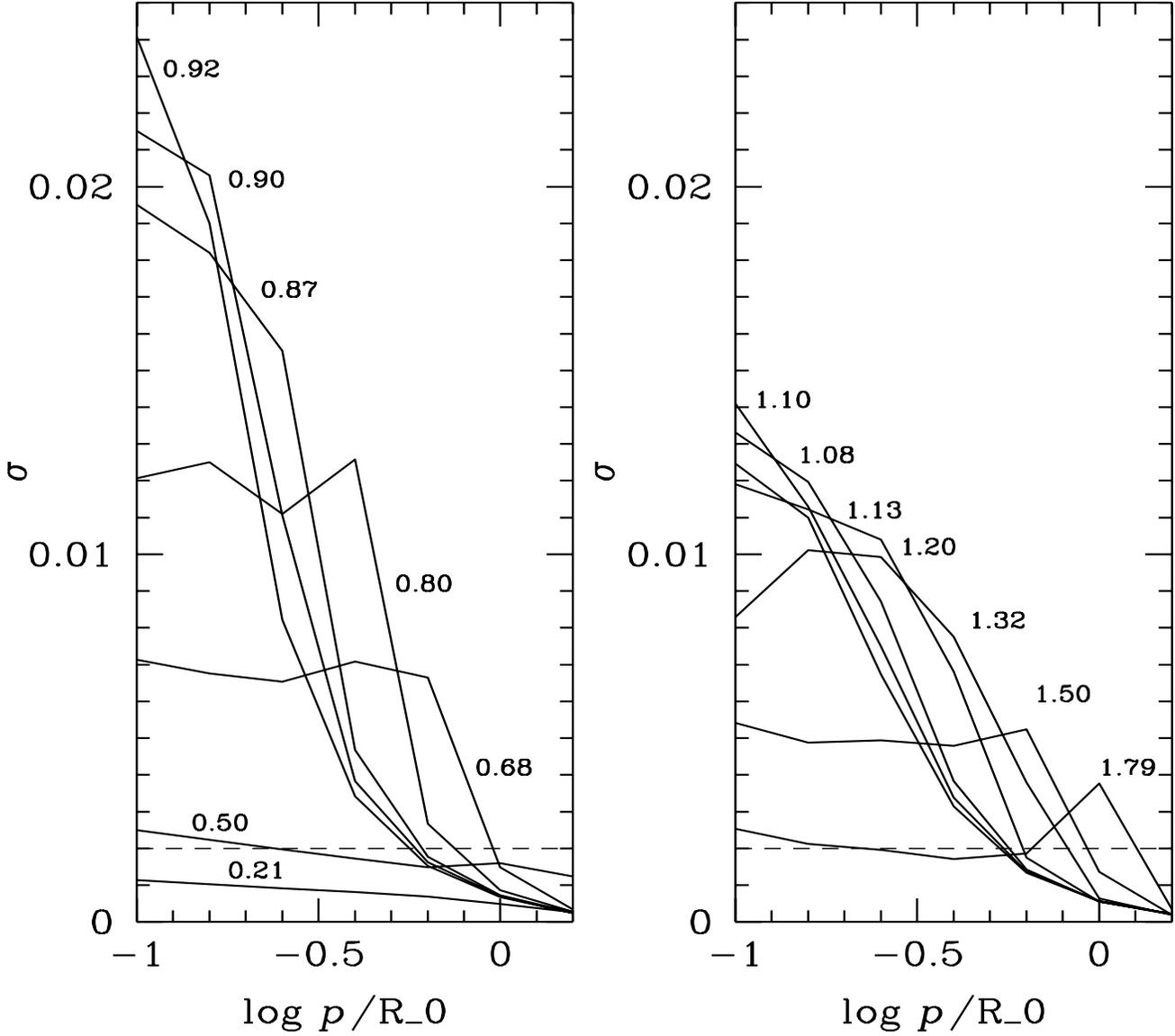

Fig. 2.— Cross-section $\sigma$ (width) for $q = 0.001$ as a function of the logarithm of the dimensionless impact parameter, parameterized by the dimensionless separation $a/R_0$. For clarity, we plot separately curves with $a/R_0 < 1$ (left) and $a/R_0 > 1$ (right); each curve is labeled with its corresponding value of $a/R_0$. The threshold $2 \times 10^{-3}$ adopted in this paper is indicated by dashed lines.